\documentclass{PoS}
\newcommand{\be}{\begin{equation}}
\newcommand{\ee}{\end{equation}}
\newcommand{\bea}{\begin{eqnarray}}
\newcommand{\eea}{\end{eqnarray}}

\newcommand{\bra}[1]{\left\langle #1 \right|}
\newcommand{\ket}[1]{\left| #1 \right\rangle}
\title{Anomalous AV$^*$V vertex function in the soft-wall holographic model of QCD}

\ShortTitle{Anomalous AV$^*$V vertex function in the soft-wall holographic model of QCD}

\author{\speaker{Fulvia De Fazio}%
         \\
        Istituto Nazionale di Fisica Nucleare, Sezione di Bari, Italy\\
        E-mail: \email{fulvia.defazio@ba.infn.it}}

\abstract{I consider the three-point function of two vector and one axial-vector currents. In the kinematic configuration in which one of the two vector currents corresponds to
an on-shell soft photon, such a vertex is described in QCD by two functions,  $w_L$ and $w_T$.
After reviewing the properties of these functions in QCD, I describe the result obtained using the soft-wall
holographic model of QCD with the addition of the Chern-Simons term and I compare it with the QCD outcome.  I also discuss a relation, proposed by
Son and Yamamoto, that connects $w_L$ and  $w_T$ to the two-point functions $\Pi_{VV}$ and $\Pi_{AA}$.}

\FullConference{Xth Quark Confinement and the Hadron Spectrum\\
		 8?12 October 2012\\
		 TUM Campus Garching, Munich, Germany}

\begin{document}

\section{Introduction}

In this paper I summarize the results obtained in \cite{Colangelo:2011xk} using a holographic approach to QCD, the soft-wall model \cite{Karch:2006pv}, for the three-point function of two vector and one axial vector currents if one of the vector currents corresponds to an on-shell soft photon.
This function enters, for example, in the calculation of the massless fermion anomalous
triangle diagrams. While the longitudinal part of such diagrams is fixed by the chiral anomaly and receives no kind of corrections, the transverse part is affected by both perturbative and non perturbative corrections to the result obtained through the computation of the leading order  diagram. On the other hand, corrections should be included
to both longitudinal and transverse parts when massive quarks are considered.

Aiming at understanding to which extent the holographic approach  reproduces known QCD features, it is interesting  to calculate these quantities in this  framework and compare them with the corresponding QCD findings.
 I also discuss  an interesting relation,  proposed  in \cite{Son:2010vc}, connecting the
transverse part of the anomalous  triangle diagrams
and the two-point left-right current correlator.

\section{Functions $w_L$ and $w_T$ in QCD}\label{sec:qcd}
I consider the correlation function of two  vector currents $J_\mu=\bar q V \gamma_\mu q$
and  an axial current  $J^5_\nu=\bar q A \gamma_\nu \gamma_5 q$, ($V$, $A$ being diagonal matrices acting on the flavour indices of the quark fields $q$),
 when one of the two vectors
corresponds to  a real, soft photon with  momentum $k\simeq 0$ and $k^2=0$:
\be
  T_{\mu \nu \sigma} (q,k)=i^2  \,\, \int d^4x \, d^4y\, e^{i\,q \cdot x-i \, k \cdot y} \bra{0} T[J_\mu(x)J_\nu^5(0)J^{em}_\sigma(y)] \ket{0}\,\,,
\label{threepoint}
\ee
with  $J^{em}_\sigma$  the electromagnetic current.
A related quantity is
 the two-point correlator  in an external electromagnetic field
$
T_{\mu \nu}(q,k)=i \,\, \int d^4x \, e^{i\,q \cdot x} \bra{0} T[J_\mu(x)J_\nu^5(0)] \ket{\gamma(k,\epsilon)}
$
since  $
 T_{\mu \nu}(q,k)= e \,\epsilon^{ \sigma} \,T_{\mu \nu \sigma}(q,k)$,
with $\epsilon^\sigma(k)$  the photon polarization vector and $e$ the electric charge unit.
For  $k\to 0$, keeping only linear terms  in $k$,   $T_{\mu \nu}$    can be written
in terms of two structure functions $w_L(q^2)$  and $w_T(q^2)$:
\be
T_{\mu \nu}(q,k)=-{i \, \over 4 \pi^2} {\rm Tr}\left[Q V A\right] \left\{ w_T(q^2)(-q^2 {\tilde f}_{\mu \nu}+q_\mu q^\lambda {\tilde f}_{\lambda \nu} -q_\nu q^\lambda {\tilde f}_{\lambda \mu})+ w_L(q^2) q_\nu q^\lambda {\tilde f}_{\lambda \mu} \right\}\,\,. \label{decomposition}
\ee
  $Q$ is the electric charge matrix and  ${\tilde f}_{\mu \nu}=\displaystyle{1 \over 2} \epsilon_{\mu \nu \alpha \beta} f^{\alpha \beta}$  the  dual   of the photon field strength $f^{\alpha \beta}=k^\alpha \epsilon^\beta -k^\beta \epsilon^\alpha$. The first term in  (\ref{decomposition}) is transverse  with respect to the axial current index, the second one  longitudinal.
The calculation of  the triangle loop diagram corresponding to  (\ref{threepoint})  when it takes contribution from a single  quark of mass $m$,  was first performed in \cite{Adler:1969gk} with the result:
\be
w_L(Q^2)=2 \, w_T(Q^2) = {2 N_c \over Q^2} \left[1 +{2 m^2 \over Q^2}\ln{m^2 \over Q^2}+{\cal O} \left({m^4 \over Q^4} \right) \right] \,\,, \label{wLT-1loop} \ee
where $Q^2=-q^2$.
Both  perturbative and nonperturbative corrections  modify this result.
Nevertheless, a non-renormalization theorem  protects $w_L$ from receiving perturbative corrections \cite{Adler:1969er}, while, in the case of $w_T$, in \cite{Vainshtein:2002nv} it has been shown that in the kinematic condition   in which one of the photons is on shell and soft $(k \to 0$),  and for $Q^2 \gg m^2$,  also $w_T$ does not receive perturbative corrections to any order. Therefore,
in the chiral limit $m=0$  one has
\be
w_L(Q^2) = {2 N_c \over Q^2}\,\,.
\label{basic-rel-0}\ee
Furthermore, discarding nonperturbative corrections, the relation holds:
\be
w_L(Q^2)=2 \, w_T(Q^2)\,\, .
\label{basic-rel}\ee

 In the chiral limit  nonperturbative corrections to $w_L$ are also absent  since    the dependence $\displaystyle w_L \propto 1/ Q^2$ reflects the contribution  of the pion pole at $Q^2=0$ to the longitudinal part of $T_{\mu\nu}$.
Nonperturbative corrections to $w_T$ exist and start at ${\cal O}\left(Q^{-6} \right)$.
For $m \neq 0$, using the   Operator Product Expansion (OPE) at large Euclidean $Q^2$, the operator ${\hat T}_{\mu \nu}$ can be written as
\be
{\hat T}_{\mu \nu}=i \,\, \int d^4x \, e^{i\,q \cdot x}  \, T[J_\mu(x)J_\nu^5(0)] =\sum_i c^i_{\mu \nu \alpha_1 \alpha_2 \dots \alpha_i}(q) \, O_i^{\alpha_1 \alpha_2 \dots \alpha_i}\,\,.,
\label{T-hat}
\ee
where  $O_i$ are local operators  and   $c^i$ coefficients computed perturbatively.
The dimension of the  $O_i$ matches the dependence of the  $c^i$ on the inverse powers of $Q^2$.
Keeping only  linear terms in the photon momentum $k$, the structure of  the OPE  for ${\hat T}_{\mu \nu}$ is
\be
{\hat T}_{\mu \nu}=\sum_i \left\{ c^i_T(q^2)(-q^2  O^i_{\mu \nu}+q_\mu q^\lambda  O^i_{\lambda \nu} -q_\nu q^\lambda  O^i_{\lambda \mu})+ c^i_L(q^2) q_\nu q^\lambda O^i_{\lambda \mu} \right\}\,\,,  \label{T-hat-ope}
\ee
so that, writing
$
\bra{0} O_i^{\alpha \beta} \ket{\gamma(k,\epsilon)}=-{i \, e  \over 4 \pi^2} \kappa_i {\tilde f}^{\alpha \beta} $,
one finds:
$w_{L,T}(Q^2)=\sum_i c^i_{L,T}(Q^2) \, \kappa_i $.
Including operators with dimension up to $D=3$, the OPE result for $w_L$ and $w_T$ reads:
\be
w_L(Q^2)=2 \, w_T(Q^2) = {2 N_c \over Q^2} \left[1 +{2 m^2 \over Q^2}\ln{m^2 \over Q^2} - {8 \pi^2 m \langle {\bar q}q \rangle \chi  \over N_c Q^2}+{\cal O} \left({m^4 \over Q^4} \right)\right] \,\, \label{wLT-2}
\ee
at large $Q^2$ (with ${\cal O}(\alpha_s)$ corrections computed in \cite{Melnikov:2006qb}).
In (\ref{wLT-2}), $\langle \bar{q}q\rangle$ denotes the vacuum quark condensate and
 $\chi$  the  so-called  magnetic susceptibility  of the  quark condensate.

As for higher order terms, the dimension $D=4$ operators can be reduced to the $D=3$ ones  using the quark equation of motion, while both $D=5$ and $D=6$ terms contribute to ${\cal O} \left( {1 \over Q^6} \right)$ order.
Remarkably, the contribution of the dimension $D=6$ operators does not vanish in the chiral limit and is responsible of the difference between $w_L$ and $2\, w_T$.
Indeed, for  $m_q=0$,   $w_L$ remains $\displaystyle w_L(Q^2)=\frac{2 N_c}{Q^2}$,  while  $w_T$, including  the leading nonperturbative correction,  reads~\cite{Knecht:2002hr,VVA-Vainshtein}:
\begin{eqnarray}
w_T(Q^2)  &=& \frac{N_c}{Q^2}\,\,
+\,\, \frac{128 \pi^3 \alpha_s \,\chi\,\langle\bar{q}q\rangle^2}{9\, Q^6}\, \,\,\,
+\,\,\,\, {\cal O}\left(\frac{1}{Q^8}\right)\, .
\label{eq.wT-OPE-mq0}
\end{eqnarray}
 $\chi$
 arises  here  factorizing  the matrix element of  four-quark operators
in the  external field  $F^{\alpha\beta}$.
%~\cite{VVA-Vainshtein}.
There might  be other ${\cal O}(1/Q^6)$ contributions in the OPE;  however, they stem from operators contributing at one loop  with small coefficients,  while the $1/Q^6$ term in~(\ref{eq.wT-OPE-mq0})
comes from  tree-level diagrams.
%~\cite{VVA-Vainshtein}.

\section{Functions $w_L$ and $w_T$ in the soft-wall AdS/QCD model}\label{sw}
The  AdS/CFT correspondence  was formulated as a duality between a type IIB string
theory defined on AdS$_5$ $\times S^5$ space  and a
${\cal N}$=4 super Yang-Mills theory with gauge group $SU(N_c)$,
for large $N_c$
\cite{Maldacena:1997re}. Subsequently, the
correspondence has been generalized as an equivalence between a
theory defined on AdS$_{d+1}\times{\cal C}$ (${\cal C}$ being a
compact manifold) and a conformal field theory defined on the flat
boundary ${\cal M}_d$ of the AdS space \cite{Witten:1998qj}; on this basis  the challenging
attempt of describing strong
interaction processes by  this approach has been undertaken. Two  ways are followed to this aim. The first one is
so-called top-down approach in which, starting from a string
theory, one tries   to derive a low-energy QCD-like theory on  ${\cal
M}_d$ through compactifications of the extra dimensions
\cite{top-down}. The second one is  the  bottom-up approach in which,
starting from $4d$ QCD,  one tries to construct its higher
dimensional dual \cite{polchinsky}.

In both cases it is necessary  to break conformal invariance,
since QCD is not a conformal theory \cite{confqcd}, and to account
for confinement.
In the bottom-up approach, conformal invariance can be broken  by allowing the
fifth  coordinate
$z$ to vary up to a maximum value $z_{max}$ of ${\cal
O}(\frac{1}{\Lambda_{QCD}})$ \cite{polchinsky,son1} (hard-wall model), or by introducing in the $5d$ AdS  space a background dilaton field
(soft-wall model) \cite{Andreev soft wall,son2}.
I consider  the soft-wall model,  defined in a five dimensional AdS space with line element
$
ds^2=g_{MN}dx^M dx^N={R^2 \over z^2}(\eta_{\mu \nu} dx^\mu dx^\nu-dz^2)$
with $M,N=0,1,2,3,5$,  $\eta_{\mu \nu}=diag(+1,-1,-1,-1)$ and $R$  is the AdS curvature radius (set to unity).
The fifth coordinate $z$ runs in the range $\epsilon \le z < + \infty$,
with $\epsilon \to 0^+$, and one introduces a background dilaton-like field
$
\Phi(z)=(cz)^2$.
With this choice,   linear Regge trajectories for light vector mesons are obtained; $c$ is a dimensionful parameter numerically fixed to $c={M_\rho \over 2}$ from  the analysis of the
spectrum of the light vector mesons. In this framework light  vector,  axial-vector  and pseudoscalar mesons can be described, with a mechanism of chiral symmetry breaking  related to
the presence of a scalar field; the light scalar meson   sector  has also been studied in \cite{Colangelo:2008us}.

To study the considered three-point function one introduces the left and right gauge fields ${\cal A}_{L\mu}^a$ and ${\cal A}_{R\mu}^a$,
dual to the $SU(N_f)_L$ and $SU(N_f)_R$ flavour currents,  ${\bar q}_L \gamma^\mu T^a q_L$ and ${\bar q}_R \gamma^\mu T^a q_R$, with $T^a$  the generators of $SU(N_f)$ \cite{Erlich:2005qh,Da Rold:2005zs,Karch:2006pv}.
 Since we want to describe  the electromagnetic  current that contains both isovector and isoscalar components, we enlarge the gauge group to $U(N_f)_L \times U(N_f)_R$   to describe its dual.
 The  gauge fields ${\cal A}_{L,R}^M$ are then combined into  a vector  $V^M=\displaystyle{{\cal A}_L^M +{\cal A}_R^M \over 2}$ and an axial-vector field  $A^M=\displaystyle{{\cal A}_L^M -{\cal A}_R^M \over 2}$, and the corresponding field strength tensors $F_{V,A}^{MN}$ are introduced:
 $F_V^{MN}=\partial^M V^N -\partial^N V^M -i[V^M,V^N]-i[A^M,A^N]$, $
 F_A^{MN}=\partial^M A^N -\partial^N A^M -i[V^M,A^N]-i[A^M,V^N] $.
 A scalar bulk field, dual to the quark bifundamental field ${\bar q}_R^\alpha q_L^\beta$, is also introduced:
$
X=X_0 e^{2i\pi}$, where
 $X_0=\displaystyle{v(z) \over 2}$ is a background field that depends only on $z$. It provides  chiral symmetry breaking, being dual to the QCD quark condensate $<{\bar q} q>$.  $\pi(x,z)$ represents the pseudoscalar meson field.
 The definition of $X$ can be further modified to $(X_0 + S) e^{2i\pi} $,  including a scalar field $S(x,z)$ describing light scalar mesons \cite{Colangelo:2008us}.
The 5d action for  the  fields ${V,A}$ and $X$ reads
 \be
 S_{YM}={1 \over k_{YM}} \int d^5x \sqrt{g}e^{-\Phi} Tr\left\{ |DX|^2-m_5^2 |X|^2 -{1 \over 2 g_5^2} (F_V^2+F_A^2) \right\} \,\,\, ,
 \label{action}
 \ee
where  $D^MX=\partial^M X-i[V^M,X]-i\{A^M,X\} $ is the covariant derivative,
$g$  the determinant of the metric tensor $g_{MN}$, $\Phi(z)$  the dilaton,  and $k_{YM}$ a parameter. Matching the two-point  function of the vector field $V$, and that of the  scalar field $S$, with the corresponding leading order perturbative QCD results
fixes  $ k_{YM}={16 \pi^2 \over N_c}$ and
 $ g_5^2={3 \over 4}$   \cite{Erlich:2005qh,Colangelo:2008us}.

To compute the functions $w_{L,T}$, following \cite{Witten:1998qj} and
 \cite{Hill:2006wu}-\cite{Brodsky:2011xx}, \cite{Son:2010vc}  a Chern-Simons  contribution to $S_{YM}$ is added:
  \be
S_{CS}({\cal A})= k_{CS}\int d^5x \,\, Tr \left[ {\cal A} F^2-{i\over 2} {\cal A}^3 F - {1\over 10} {\cal A}^5 \right]. \label{chernsimons}
 \ee
Terms in  the Chern-Simons action $S_{CS}$ proportional to higher odd powers of ${\cal A}_{L,R}$ do not contribute to the  $AV^*V$ vertex considered  here, so that they can be neglected, keeping in (\ref{chernsimons}) only  the terms
 $\displaystyle Tr \left[ {\cal A}_{L,R} F^2_{L,R}\right]=\epsilon_{ABCDE}\, Tr \left[{\cal A}^A_{L,R} F_{L,R}^{BC} F_{L,R}^{DE} \right]$, with  $A,  \dots, E$  indices of the $5d$ coordinates.
 Moreover, since  the Chern-Simons actions  are invariant only up to a boundary term, in \cite{Colangelo:2011xk} a boundary term has been included to make explicit the invariance under a
 vector gauge transformation, obtaining:
 $
 S_{CS+b} = 3\,  k_{CS} \,\,  \epsilon_{ABCDE} \,  \int d^5x \,\,Tr \left[{A}^A \left\{F_{V}^{BC}, F_{V}^{DE} \right \} \right]  $.
 The constant $k_{CS}$ is fixed below.
 
 Our starting point   is  the  effective action
$S_{5d}^{eff}=S_{YM}+S_{CS+b}$.
Exploiting  the AdS/QCD correspondence, the 5d action is dual to  the QCD generating functional relative to a given operator $O(x)$ provided that the source of $O(x)$ coincides with the $z=0$ boundary value, $f_0(x)=f(x,0)$, of the dual field $f(x,z)$ in the 5d action:
$\biggl\langle e^{i\int d^4 x\;0(x )\,f_0(x)}\biggr\rangle_{QCD}=
  e^{iS_{5d}^{eff}[f(x,z)]}$.
According to this prescription, the  functions  $w_L$ and $w_T$   are computed   by  a functional derivation of the 5d action.
One has first to define ${\tilde G}^a_\mu(q,z)$ as the Fourier transform with respect to the 4d coordinates $x^\mu$ of a generic gauge field $G^a(x,z)=V^a(x,z)$ and $A^a(x,z)$
($a$ flavour index) and then introduce the bulk-to-boundary propagator $G(q,z)$ as: ${\tilde G}^a_\mu(q,z)=G(q,z)G_{\mu 0}^a(q)$,  where $G_{\mu0}^a(q)$ is the source field.
In the case of the vector and axial-vector fields of momentum $q$ we consider two projectors
$
P_{\mu \nu}^\perp = \eta_{\mu \nu}-{q_\mu q_\nu \over q^2}$,
$P_{\mu \nu}^\parallel = {q_\mu q_\nu \over q^2}$,
in such a way that the vector and axial-vector bulk-to boundary propagators  are expressed in terms of  the transverse and longitudinal parts:
\be
{\tilde V}^a_\mu(q,z)=V_\perp(q,z)P_{\mu \nu}^\perp V_{ 0}^{a\nu }(q)\hskip 0.3cm , \hskip 0.3 cm
{\tilde A}^a_\mu(q,z)=A_\perp(q,z)P_{\mu \nu}^\perp A_{0}^{a\nu}(q)+A_\parallel(q,z)P_{\mu \nu}^\parallel A_{ 0}^{a\nu}(q) \,\,,\label{perp-par} \ee
imposing as boundary conditions $V_\perp(q,0)=1$ and $A_\perp(q,0)=A_\parallel(q,0)=1$ (the behaviour at   $z \to \infty$ is discussed later) and accounting for the fact that  the (conserved) vector field is transverse.
The longitudinal component of ${\tilde A}$ is written as ${\tilde A}^{a \parallel}_\mu(q,z)=A_\parallel(q,z)P_{\mu \nu}^\parallel A_{\nu 0}^a(q)=i\,q_\mu {\tilde \phi}^a$.

From the 5d action
  a set of equations of motion are obtained in the gauge $V_z=A_z=0$:
\bea
&&\partial_y \left({e^{-y^2} \over y}  \,\, \partial_y V_\perp \right)-{\tilde Q}^2 {e^{-y^2} \over y}V_\perp=0 \label{EOM-V} \\
&&\partial_y \left({e^{-y^2} \over y}  \,\, \partial_y A_\perp  \right)-{\tilde Q}^2 {e^{-y^2} \over y}A_\perp -{g_5^2 v^2(y) e^{-y^2} \over y^3}A_\perp=0 \label{EOM-A} \\
&&\partial_y \left({e^{-y^2} \over y} \,\, \partial_y  {\tilde \phi}^a  \right)+{g_5^2 v^2(y) e^{-y^2} \over y^3}  ({\tilde \pi}^a-{\tilde \phi}^a)=0 \label{EOM-phi-pi-1} \\
&&{\tilde Q}^2 (\partial_y  {\tilde \phi}^a) +{g_5^2 v^2(y) \over y^2}\partial_y {\tilde \pi}^a=0 \label{EOM-phi-pi-2} \,\,\, \eea
where $y=cz$ and ${\tilde Q}^2=\displaystyle{Q^2 \over c^2}$, with  $Q^2=-q^2>0$ (the  notation $V=V_\perp$, $A=A_\perp$ has been adopted).  From the relation
$ {\tilde \phi}^a(q,y)=-i{q^\mu \over q^2}A_\parallel(q,y) P_{\mu \nu}^\parallel A_{\nu 0}^a(q) $
and writing $\displaystyle {\tilde \pi}^a(q,y)=-i{q^\mu \over q^2}\pi(q,y) A_{\mu 0}^a(q)$, it turns out that  $\pi(q,y)$ and $A_\parallel(q,y)$ obey the same equations (\ref{EOM-phi-pi-1}) and (\ref{EOM-phi-pi-2})
as $\tilde \pi^a$ and $\tilde \phi^a$.

An equation can also be derived for the field $X_0 = \frac{1}{2} \, v$:
$
\partial_y \left({e^{-y^2} \over y^3}  \,\, \partial_y v(y)  \right)+{3 e^{-y^2} \over y^5} v(y)=0  $
the regular solution of which reads
$
 v(y)\sim \Gamma \left({3 \over 2}\right) \, y \, U\left({1\over 2}, 0, y^2\right)$ and can be
 expanded for $y \to 0$:   $v(y)\to C_1 y + C_2 y^3$. On the basis of the holographic dictionary \cite{Erlich:2005qh}, one argues that the quark mass,
 responsible of explicit chiral  symmetry breaking, enters in the coefficient $C_1$, and the quark condensate, governing the spontaneous chiral symmetry breaking,  enters in  $C_2$:
 $ m_q \propto C_1$, $ \sigma \propto \langle {\bar q} q\rangle \propto C_2 $.
 However, since
  $C_1$ and $C_2$ are related,  a proportionality  between  $m_q$ and  $\langle {\bar q} q\rangle$ is implied, which does not hold in QCD.
   This  feature of the soft-wall model  could be corrected adding potential terms
$V(|X|)$ to the action. However, in the following  the form
$v(y)=\frac{m_q}{c} \, y + \frac{\sigma}{c^3} \, y^3$ is assumed \cite{lebed}.

Determining   $w_L$ and $w_T$, by the AdS/CFT prescription requires  a functional derivation of the effective 5d action.
The Chern-Simons action is written as
$
S_{CS+b}=48 \, k_{CS}  \, d^{ab} \, {\tilde F}^{\mu \nu}_{em} \int d^5x \, A^b_\nu \, \partial_z V^a_\mu $,
with $d^{ab}={1 \over 2}Tr[Q\{T^a,T^b\}]$, and ${\tilde F}^{\mu \nu}_{em}$  the external photon field strength. Performing the functional derivation one has
\be
d^{ab} (2\pi)^{-4} \delta^4(q_1+q_2) \langle J_\mu^{V} J_\nu^{A} \rangle_{\tilde F}^{\perp \perp(\parallel)}=
{\delta^2 S_{CS+b} \over \delta V_{\mu0}^{a \perp}(q_1) \, \delta A_{\nu0}^{b \perp(\parallel)}(q_2) } \label{funct-der} \,\,.
\ee
On the other hand,
the correlation function of a vector  and an axial vector current  in the external electromagnetic background field reads, in terms of s $w_{L}$ and $w_{T}$,
\be
d^{ab} \langle J_\mu^{V} J_\nu^{A} \rangle_{\tilde F}\equiv  i \int d^4 x \, e^{i q x} \langle T \{J_\mu^{Va} (x) J_\nu^{Ab}(0) \} \rangle_{\tilde F}
=  d^{ab} {Q^2 \over 4 \pi^2} P_{\mu \alpha}^\perp \left[P_{\nu \beta}^\perp w_T(Q^2) + P_{\nu \beta}^\parallel w_L(Q^2) \right] \, {\tilde F}^{\alpha \beta} \,\,,
\label{JJ-F}
\ee
so that  comparison of  (\ref{JJ-F}) with (\ref{funct-der}) gives
\be
w_L(Q^2)=-{2 N_c \over Q^2}\int_0^{\infty} dyA_\parallel(Q^2,y)  \partial_y V(Q^2,y) \hskip 0.25cm , \hskip 0.25cm
w_T(Q^2)=-{2 N_c \over Q^2}\int_0^{\infty} dy A_\perp(Q^2,y) \partial_y V(Q^2,y) \label{wLT-ris}\,\,.
\ee
The choice $\displaystyle k_{CS}=-{N_c \over 96 \, \pi^2}$  reproduces the leading term in the QCD OPE (\ref{basic-rel}).

To compare these results with the QCD findings one needs to solve the equations of motion for  $V$, $A_\perp$, $A_\parallel$.
Eq.(\ref{EOM-V})  for $V(Q^2,y)$ can be exactly solved with the boundary conditions $V(Q^2,0)=1$   and $V(Q^2,\infty)=0$,  with the result
$
V(Q^2,y)=\Gamma \left( 1+ {{ Q}^2 \over 4 c^2} \right) U\left({ { Q}^2 \over 4 c^2},0,y^2 \right)$,
where  $U$ is the Tricomi confluent hypergeometric function.
The calculation is more difficult for $A_\perp$ and $A_\parallel$ since Eqs.(\ref{EOM-A}) and (\ref{EOM-phi-pi-2})  involve   the  function $v(y)$.
Expanding  in the inverse powers of $Q^2$, the   result is found \cite{Colangelo:2011xk}:
\be
w_T(Q^2)={ N_c \over Q^2}\left(1-{g_5^2 m_q^2 \over 3 Q^2}-{2g_5^2 m_q^2 c^2 \over 5 Q^4}+
{g_5^4 m_q^4 \over 6 Q^4}-{8g_5^2 m_q \sigma \over 5 Q^4}  \right) + {\cal O}\left({1 \over Q^8}\right) \,,
\label{wT-final}
\ee
\begin{equation}\label{wLcompl}
 w_L(Q^2) = \frac{2N_c}{Q^2} - \left[1 - \pi(Q^2,0)\right]\,N_c\left[\frac{g_5^2\, m_q^2}{Q^4} +\frac{4g_5^2\, m_q \, \sigma}{Q^6} - \frac{2g_5^4\,m_q^4}{3Q^6} + {\cal O}\left(\frac{1}{Q^8}\right)\right]\,,
\end{equation}
where the boundary condition of the chiral field $\pi(Q^2,0)$ appears in the expression of $w_L$.
For $m_q=0$, at leading order in the $1/Q^2$ expansion, the QCD results in  (\ref{basic-rel-0}-\ref{basic-rel}) are recovered.
The result for $w_L$ holds even for $\sigma \neq 0$, while that for $w_T$ is modified. Considering Eq. (\ref{wT-final}) for $m_q=0$ and $\sigma \neq 0$, one can observe that the first correction to the leading term is of ${\cal O}({1 \over Q^8})$.  On the other hand,  Eq. (\ref{eq.wT-OPE-mq0}), obtained in QCD in the same limit, shows the first correction  of ${\cal O}({1 \over Q^6})$. This could be an indication that in the considered holographic model the susceptibility of the chiral condensate $\chi$ vanishes. Furthermore, a mismatch is found also comparing the solutions in the most general case $m_q \neq 0$, $\sigma \neq 0$ since the logarithmic term in the QCD result  (\ref{wLT-2}) is missing in the soft-wall model result.

\section{Two-point functions and  Son-Yamamoto relation}
Let us consider the left-right    correlator $\Pi_{LR}= \Pi^{VV}_\perp-\Pi^{AA}_\perp$, where $\Pi^{VV}_\perp$ and $\Pi^{AA}_\perp$ are the transverse invariant structures that appear in the two-point functions of
  the vector and axial-vector currents $J^{a}_\mu=\bar{q}\gamma_\mu T^a q$
and axial-vector currents   $J^{5a}_\mu=\bar{q}\gamma_\mu \gamma_5 T^a q$
\be
\Pi_{\mu\nu}^{ab}(q)\,=   \,\,i\, \int d^4x\,\,e^{i q x}\,\,\langle0| T\{\, J^a_\mu (x) \, J^b_\nu(0)\}|0\rangle
=\, (q_\mu q_\nu -q^2 g_{\mu\nu})\, \delta^{ab} \, \Pi_\perp  (q^2)
\,+\, q_\mu q_\nu \, \delta^{ab} \, \Pi_\parallel(q^2)   \,\,\, .\label{PI_LR}
\ee
In \cite{Son:2010vc} a relation has been derived that links this quantity to the pion decay constant $F_\pi$  and the structure function $w_T$ in massless QCD and for any positive and negative $Q^2$:
\be
w_T(Q^2)=\frac{N_c}{Q^2}+\frac{N_c}{F_\pi^2}Ê\, \Pi_{LR}(Q^2) \,\,\,\, . \label{son-yam}
\ee
Computing $\Pi_{LR}$ in the soft wall model  and  expanding  in the inverse powers of $ Q^2$  one gets \cite{Colangelo:2011xk}
\begin{equation}
 \Pi_{LR}(Q^2)\,  = \, -\, \frac{N_c\, \sigma^2}{10\pi^2\, Q^6} + {\cal O}\left(\frac{1}{Q^8}\right)\, . \label{PILR-res}
\end{equation}
The first contribution to  $\Pi_{LR}$ is of  $\displaystyle {\cal O}\left(1/Q^{6}\right)$, and has the same negative sign as the corresponding  dimension six condensate
 in QCD \cite{SVZ,PI-Narison}.
On the other hand, for  $m_q=0$,  the  leading power correction to $w_T$  is
$\displaystyle w_T(Q^2)=\frac{N_c}{Q^2}  \, \left(1\, + {\cal O}\big(\frac{1}{Q^{6}}\big)\right)$, so that  the $Q^2$  dependencies  of the two sides of  (\ref{son-yam}) do not match, similarly to what  is found in  \cite{Son:2010vc}. Other discussions about this relation can be found in  and in  other holographic models \cite{Colangelo:2012ip,Knecht:2011wh,Iatrakis:2011ht,Gorsky:2012ui}.

\section{Conclusions}
The functions $w_L$ and $w_T$ can be determined in the
 soft-wall model  with the Chern-Simons term in the action, providing the results (\ref{wLT-ris}) .
In the chiral  limit,  the result  (\ref{basic-rel-0}) for $w_L$ is recovered and holds  also for non vanishing chiral condensate.  As for $w_T$,
when $m_q=0$ and $\sigma=0$  the described calculation reproduces the QCD result and the relation (\ref{basic-rel}).

Away from the chiral limit,  mismatches are found in the $1/Q^2$ expansion when comparing the holographic results for $w_{L,T}$ to the QCD outcome, a consequence of the choice
 of the simplest inclusion of the quark mass in the holographic framework through the expression of $v(y)$.
In particular,
in the  expansion of $w_T$ for large  $Q^2$,  the next-to-leading contribution in $w_T$ is ${\cal O}(1/Q^8)$ in the holographic model, while it is ${\cal O}(1/Q^6)$ in QCD.
Since in QCD this correction involves the magnetic susceptibility  $\chi$ of the quark condensate, a simple interpretation of this result could be  that, in the soft-wall  model, $\chi$ vanishes.
Another explanation could be that
 operators like  $O_{\mu \nu}=\bar q\sigma_{\mu\nu} q$  should be included in the dual approach \cite{Domokos:2011dn}.
Finally, this calculation eveils a violation of  a proposed relation between $w_T$ and $\Pi_{LR}$, Eq.(\ref{son-yam}).

\acknowledgments
I thank P. Colangelo, F. Giannuzzi, S. Nicotri and J.J. Sanz-Cillero for collaboration.

\end{document}